\documentclass[cameraready]{Interspeech}
\usepackage{multirow}

\title{Continual Adaptation for Pacific Indigenous Speech Recognition}

\author[affiliation={1}]{Yang}{Xiao}
\author[affiliation={1}]{Aso}{Mahmudi}
\author[affiliation={1}]{Nick}{Thieberger}
\author[affiliation={2}]{Eliathamby}{Ambikairajah}
\author[affiliation={1}]{Eun-Jung}{Holden}
\author[affiliation={1}]{Ting}{Dang}

\address{
    $^1$ The University of Melbourne \\
    $^2$ UNSW Sydney
}

\email{yxiao9550@student.unimelb.edu.au, ting.dang@unimelb.edu.au}
\keywords{Speech recognition, Pacific languages}

\usepackage{comment}
\usepackage{soul, xcolor}

\definecolor{myblue}{RGB}{200,220,235}

\begin{document}

\maketitle

\begin{abstract}
Speech foundation models struggle with low-resource Pacific Indigenous languages because of severe data scarcity. Furthermore, full fine-tuning risks catastrophic forgetting. To address this gap, we present an empirical study adapting models to real-world Pacific datasets. 
We investigate the impact of data volume, adaptation strategies, and representational drift on speech foundation models for various Pacific languages. Additionally, we analyze a continual learning framework for sequential language acquisition. Empirical results across three distinct Pacific Indigenous languages demonstrate that adapting to these linguistically distant languages induces severe internal representational drift.
Consequently, these models face a strict plasticity and stability dilemma. While LoRA adapts well initially, it suffers from catastrophic forgetting during sequential learning. Ultimately, this study highlights the urgent need for robust adaptation strategies tailored to underrepresented languages.
\end{abstract}

\section{Introduction}


Despite rapid progress in automatic speech recognition (ASR)~\cite{besacier2014automatic,yadav2022survey}, the vast majority of the world’s languages remain excluded from modern speech technologies~\cite{whisper}. This disparity is particularly pronounced for Pacific languages~\cite{Australian1,Australian2,Australian3}, many of which are spoken by relatively small communities, exhibit limited standardized orthographies, and lack large-scale annotated corpora. Moreover, these languages are not merely low-resource, they are often distributionally distant from the high-resource languages that dominate contemporary speech datasets~\cite{xiao2025dg,wu2021cross}. As speech interfaces become embedded in education, public services, and digital communication, the inability of current systems to robustly support such languages represents both a scientific and societal limitation~\cite{xiao2025adakws}.

Speech foundation models (SFMs)~\cite{shi2024ml} have been proposed as a scalable solution to low-resource ASR~\cite{oasr,conneau2020unsupervised}. Large self-supervised models pretrained on multilingual audio, such as wav2vec 2.0~\cite{baevski2020wav2vec} and Whisper~\cite{whisper}, learn acoustic representations that can be adapted to new languages with limited supervision. Empirically, these models demonstrate strong transfer performance across a range of languages, fueling the assumption that pretrained multilingual speech representations are largely language-agnostic and universally adaptable~\cite{walworth2021multilingualism}. 

However, this assumption has been validated predominantly on languages that are typologically well documented and often structurally aligned with pretraining corpora~\cite{sadhu2020continual,liu2024exploration,cl-masr,xiao2026adapting}. The extent to which SFMs generalize to linguistically divergent and underrepresented language families, such as many Pacific languages, remains insufficiently examined. Crucially, while incremental refinement paradigms have proven highly effective for standard audio and speech tasks~\cite{xiao2022rainbow, peng2024dark},  adaptation in this setting may not constitute incremental refinement of shared representations. Instead, it may require substantial internal restructuring to accommodate phonological inventories, syllable structures, and prosodic systems that are weakly represented, or entirely absent, in pretraining data.

This observation motivates a shift in perspective. Rather than asking whether SFMs can achieve acceptable recognition accuracy on Pacific languages, we ask a more fundamental question:
\emph{Does adaptation to linguistically distant, low-resource languages preserve representational stability, or does it induce large-scale representational drift and catastrophic forgetting?}
This distinction is critical. If fine-tuning behaves as smooth transfer, then pretrained representations act as a stable substrate onto which new languages can be efficiently mapped. If, however, adaptation entails substantial representational reorganization, then the scalability of current SFM paradigms to underrepresented languages is fundamentally constrained by a plasticity–stability trade-off~\cite{ewc,shi2025clep}.

Prior low-resource ASR studies largely evaluate final performance metrics (e.g., word error rate), implicitly treating adaptation as a black-box optimization problem~\cite{liu2024exploration,kwok2025two,xiao2022rainbow,xiao2022continual}. Such evaluation obscures the internal dynamics of representation change and provides limited insight into whether learned features remain stable across sequential or cross-lingual adaptation. In settings characterized by distributional mismatch, such as Pacific language transfer, this omission becomes consequential. Without analyzing representational drift and forgetting, it is impossible to determine whether performance gains reflect efficient reuse of pretrained structure or extensive rewriting of model parameters.

In this work, we reconceptualize adaptation to underrepresented Pacific languages as a stress test of representation stability in speech foundation models. We systematically analyze how standard fine-tuning behaves when transferring to languages that are both low-resource and structurally underrepresented in pretraining data. Specifically, we: i) quantify cross-lingual transfer effectiveness under limited supervision 
; ii) measure representational drift during adaptation to characterize internal restructuring 
; iii) evaluate catastrophic forgetting effects to assess the stability of previously learned capabilities and the resulting forgetting. We collected data from three different Pacific languages, and our empirical findings show that linguistic distance forces internal representation drift, which triggers catastrophic forgetting. Furthermore, these results reveal that current methods cannot resolve the forgetting during continual adaptation. By centering underrepresented Pacific languages, this study is essential for developing adaptation strategies that are not only data-efficient but also structurally robust to linguistic diversity.

\section{Pacific Indigenous Speech Corpus}
To thoroughly evaluate our adaptation strategies under real-world conditions, we introduce a newly curated speech corpus comprising three underrepresented Pacific Indigenous languages. This dataset encompasses \emph{Bislama}~\cite{thieberger2023bislama}, \emph{Nafsan}~\cite{PARADISEC_NT1_2025}, and \emph{Lelepa}~\cite{lacrampe2017lelepa}. This selection offers diverse linguistic features and varying resource levels. Specifically, Bislama is an English-based creole and the national common language. It combines English vocabulary with Melanesian grammar. In contrast, Nafsan and Lelepa are Indigenous Pacific languages in the Austronesian family. While they continue to be spoken everyday, they lack text resources. Data for this study is curated by the Pacific and Regional Archive for Digital Sources in Endangered Cultures (PARADISEC). Including both a common creole and isolated Indigenous languages provides a strict contrast to evaluate model adaptation.


As detailed in Table \ref{tab:locale_duration_stats}, the corpus contains 23,843 audio samples. This collection yields about 32.13 hours of transcribed speech. The data distribution clearly shows the severe lack of resources common in endangered languages. Specifically, Nafsan and Bislama form the larger subsets with 14.83 and 13.75 hours. In contrast, Lelepa present extreme low-resource scenarios, providing only 3.55 for modeling. Furthermore, the average sample lasts 4.85 seconds. This duration provides a stable time frame for training the acoustic models.  


\begin{table}[t]
\caption{Sample duration statistics across different locales. Specifically, the terms shortest, longest, and average indicate the sample durations in seconds.}
\vspace{-3mm}
\label{tab:locale_duration_stats}
\centering
\setlength{\tabcolsep}{4.5pt}
\scalebox{0.7}{
\begin{tabular}{l c c c c c}
\toprule
\textbf{Locale} & \textbf{Samples} & \textbf{\shortstack{Total \ Duration (H)}} & \textbf{\shortstack{Shortest (S)}} & \textbf{\shortstack{Longest (S)}} & \textbf{\shortstack{Average (S)}} \\
\midrule
Bislama & 11,576 & 13.75 & 0.17 & 34.65 & 4.28 \\
Nafsan  & 9,245  & 14.83 & 0.48 & 67.55 & 5.78 \\
Lelepa  & 3,022  & 3.55  & 0.31 & 24.06 & 4.23 \\

\midrule
\textbf{Total} & \textbf{23,843} & \textbf{32.13} & \textbf{0.17} & \textbf{67.55} & \textbf{4.85} \\

\bottomrule
\end{tabular}}
\vspace{-6mm}
\end{table}

\section{Methodology and Experimental Setup}
\subsection{Cross-lingual Transfer}
\label{subsec:cross_lingual_transfer}
To analyse the effectiveness of cross-lingual transfer, we adapt a multilingual speech foundation model pretrained on high-resource languages to each Pacific language in our corpus. Specifically, we fine-tune the Whisper-Small~\cite{whisper} model on Bislama, Nafsan, and Lelepa with progressively increasing amounts of data, while keeping the remaining audio reserved for validation and testing. For each supervision level, we compare two adaptation strategies: (i) full fine-tuning of all model parameters and (ii) parameter-efficient adaptation using LoRA of the encoder and decoder~\cite{hu2022lora}. Cross-lingual transfer effectiveness is quantified using character error rate (CER) and word error rate (WER) on the held-out test sets of each language. 

\subsection{Representational Drift Analysis}
\label{subsec:rep_drift}
To characterise how adaptation restructures internal representations, we measure representational drift in the Whisper-Small model before and after fine-tuning on each Pacific language. For a fixed evaluation set comprising balanced utterances from Bislama, Nafsan, and Lelepa, we extract hidden states from all encoder and decoder layers using (i) the original pretrained checkpoint and (ii) the adapted model under each fine-tuning configuration (full and LoRA).

For each layer, we compute the cosine distance between pre- and post-adaptation activations for every utterance and then average across the evaluation set to obtain a layer-wise drift magnitude. We subsequently apply min–max normalisation per language for comparison~\cite{pasad2021layer,pasad2023comparative}.

We compare normalised drift curves separately for encoder and decoder layers to reveal which component bears the bulk of structural change, and we contrast profiles across full fine-tuning and LoRA. This analysis links external performance gains to internal dynamics, distinguishing settings where cross-lingual transfer behaves as smooth refinement of pretrained features from those where successful adaptation requires substantial representational reorganisation and thus poses a higher risk of catastrophic forgetting in the Section~\ref{sec:results}.

\subsection{Continual Learning Analysis}
\label{subsec:continual_learning}
To evaluate the stability–plasticity trade-off when adapting to multiple underrepresented languages, we cast our experiments into a continual learning setting~\cite{de2021continual}. Starting from the same pretrained Whisper-Small checkpoint, the model sequentially learns pairs or sequences of Pacific languages (e.g., Nafsan followed by Lelepa), reflecting increasing linguistic divergence and data scarcity. For each language in the sequence, we fine-tune the model under the same adaptation strategies as in Sections~3.1 and~3.2 (full fine-tuning and LoRA) and save a checkpoint at the end of each language-specific training phase.

After completing training on a new target language, we evaluate the model on the test sets of both the newly learned language and the previously seen ones, as well as on the original high-resource reference language (English). This protocol directly exposes catastrophic forgetting as the degradation in English and earlier Pacific languages after adaptation to the latest low-resource target. The sequential Nafsan$\rightarrow$Lelepa results, summarised in Table~5, illustrate how full fine-tuning and LoRA differ in their trade-off between immediate target accuracy and retention of prior capabilities. We conduct these sequential experiments separately for each adaptation strategy, using the same data splits and optimisation hyperparameters as in the cross-lingual transfer setup. By comparing pre- and post-adaptation error rates across languages (Tables~3–5), we can identify configurations—such as encoder-focused, parameter-efficient updates—that improve robustness to catastrophic forgetting while maintaining competitive performance on newly acquired Pacific languages.

\subsection{Experimental Setup}
All experiments use the Whisper-Small~\cite{whisper} architecture as the base speech foundation model. We follow the corpus definition in Section~2 and split each language into 80\% training, 10\% validation, and 10\% test, stratified by speaker where possible to reduce speaker overlap. Audio is resampled to 16~kHz mono and normalised using the standard preprocessing pipeline from the Whisper implementation.

Fine-tuning is performed with AdamW, a peak learning rate of $1\times10^{-4}$, a batch size of 16, a linear warm-up over the first 500 steps, and a cosine decay schedule thereafter. To support the target languages, we extend the original Whisper vocabulary with characters observed in Bislama, Nafsan, and Lelepa transcripts; new token embeddings are initialised with the average of the pretrained vocabulary, while all other parameters are loaded from the multilingual checkpoint. We apply mild data augmentation via random time-shifts and additive in-domain background noise to keep the signal-to-noise ratio within a realistic conversational range. For every combination of language, budget, and adaptation strategy (full fine-tuning or LoRA), we train three runs with different random seeds and report the averaged CER and WER to ensure a reliable evaluation.

Representational drift analysis uses a separate held-out evaluation subset, disjoint from both training and test splits. For each trained checkpoint, we pass this subset through the model in teacher-forced mode, cache the hidden states of every encoder and decoder layer, and compute cosine-distance–based drift per layer with min–max normalisation per language.

In the continual learning experiments, after each language-specific training phase, we evaluate on all previously seen Pacific languages and on English to quantify catastrophic forgetting. Unless otherwise noted, model selection and early stopping in all settings are based on validation CER for the current target language.

\begin{table}[t]
\centering
\caption{Comparison of Fine-tuning Performance: Full FT vs. LoRA (Encoder-Decoder) Across Training Hours}
\vspace{-3mm}
\label{tab:comparison_encdec}
\resizebox{0.8\linewidth}{!}{
\begin{tabular}{lcccc}
\toprule
\textbf{Language} & \textbf{Hours} & \textbf{Setting} & \textbf{Test CER} & \textbf{Test WER} \\ 
\midrule
\multirow{10}{*}{Bislama} & 0.5 & Full FT & \textbf{18.88} & \textbf{41.97} \\
                     & 0.5 & LoRA   & 20.49 & 45.88 \\ \cmidrule{2-5}
                     & 1.0 & Full FT & \textbf{14.57} & \textbf{32.95} \\
                     & 1.0 & LoRA   & 17.60 & 39.51 \\ \cmidrule{2-5}
                     & 2.0 & Full FT & \textbf{15.05} & \textbf{30.92} \\
                     & 2.0 & LoRA   & 15.70 & 33.94 \\ \cmidrule{2-5}
                     & 5.0 & Full FT & \textbf{10.28} & \textbf{22.61} \\
                     & 5.0 & LoRA   & 11.77 & 25.55 \\ \cmidrule{2-5}
                     & 10.0 & Full FT & \textbf{9.30} & \textbf{19.64} \\
                     & 10.0 & LoRA   & 10.57 & 22.62 \\ 
\midrule
\multirow{10}{*}{Nafsan} &  0.5 & Full FT & 28.28 & 71.96 \\
                     & 0.5 & LoRA   & \textbf{36.35} & \textbf{88.08} \\ \cmidrule{2-5}
                     & 1.0 & Full FT & \textbf{28.63} & \textbf{69.33} \\
                     & 1.0 & LoRA   & 28.70 & 73.25 \\ \cmidrule{2-5}
                     & 2.0 & Full FT & \textbf{31.00} & \textbf{72.18} \\
                     & 2.0 & LoRA   & 36.86 & 81.40 \\ \cmidrule{2-5}
                     & 5.0 & Full FT & \textbf{20.63} & \textbf{52.05} \\
                     & 5.0 & LoRA   & 23.79 & 58.95 \\ \cmidrule{2-5} & 10.0 & Full FT & \textbf{18.83} & \textbf{47.84} \\
                     & 10.0 & LoRA   & 19.51 & 50.75 \\ 

\midrule
\multirow{6}{*}{Lelepa} &  0.5 & Full FT & \textbf{31.74} & \textbf{79.16} \\
                     & 0.5 & LoRA   & 44.19 & 96.50 \\ \cmidrule{2-5}
                     & 1.0 & Full FT & \textbf{38.17} & \textbf{84.23} \\
                     & 1.0 & LoRA   & 45.29 & 92.48 \\ \cmidrule{2-5}
                     & 2.0 & Full FT & 39.36 & 84.10 \\
                     & 2.0 & LoRA   & \textbf{33.43} & \textbf{75.66} \\ 
\bottomrule
\end{tabular}}
\vspace{-6mm}
\end{table}

\section{Results and Analysis}
\label{sec:results}


\subsection{Cross-Lingual Adaptation Effectiveness}

To quantify cross-lingual transfer under limited supervision, we evaluate the model performance across different training durations and fine-tuning methods. Table \ref{tab:comparison_encdec} details the error rates for Bislama, Nafsan, and Lelepa. First, the results reveal a direct relationship between data volume and adaptation success. As the training time increases, the error rates generally decrease. For instance, using full fine-tuning, Bislama reaches its lowest WER of 19.64 at 10.0 hours.

However, the transfer effectiveness differs significantly across the languages. Bislama shows rapid improvements even with only 1.0 hour of training data. This quick adaptation likely happens because Bislama shares many linguistic features with English~\cite{thieberger2023bislama}. In contrast, Nafsan shows limited progress under extremely low-resource conditions. Between 0.5 and 2.0 hours, the performance of Nafsan remains unstable. A noticeable improvement for Nafsan only occurs when the data volume reaches 5.0 hours.

\begin{figure}[t]
    \centering
    \includegraphics[width=0.8\linewidth]{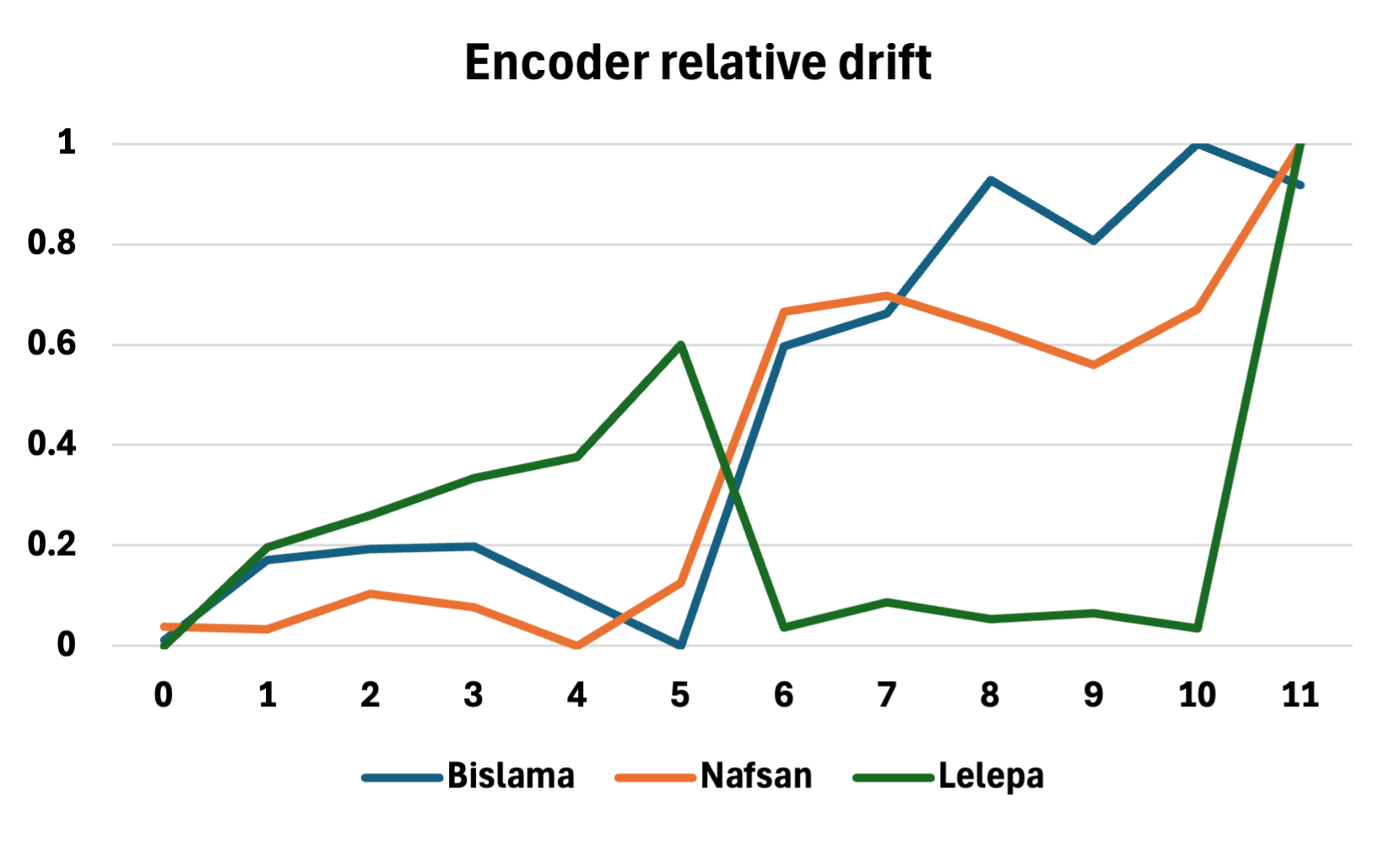}
    \vspace{-6mm}
    \caption{Relative representational drift across the twelve encoder layers.}
    \label{fig:encoder}
    \vspace{-4mm}
\end{figure}

\begin{figure}[t]
    \centering
    \includegraphics[width=0.8\linewidth]{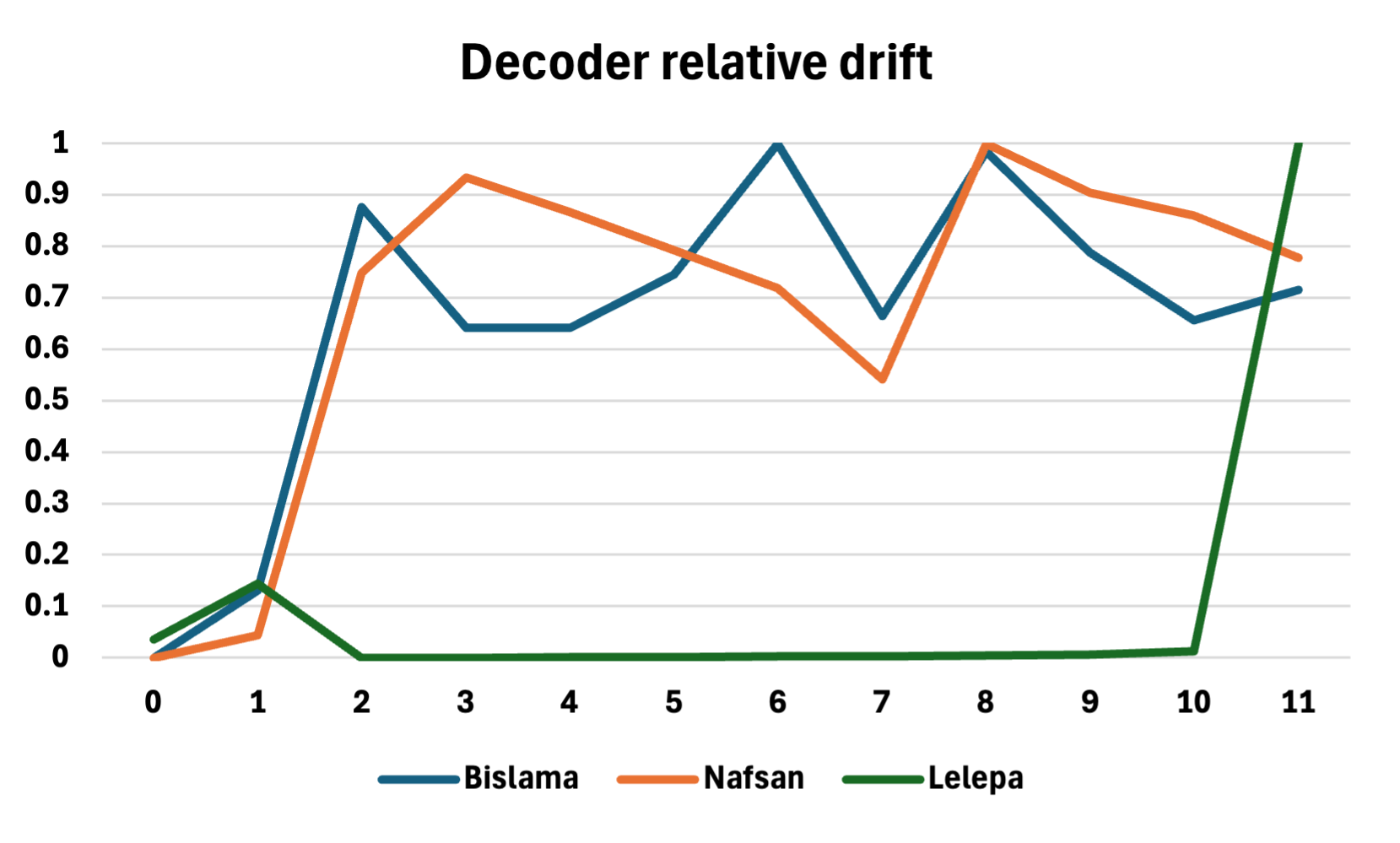}
    \vspace{-4mm}
    \caption{Relative representational drift across the twelve decoder layers.}
    \label{fig:decoder}
    \vspace{-6mm}
\end{figure}

Furthermore, the table highlights a critical performance shift between the two adaptation strategies. For Bislama and Nafsan, full fine-tuning consistently achieves lower error rates than Low-Rank Adaptation across all time settings. Conversely, Lelepa presents a unique scenario under extreme data scarcity. At 0.5 and 1.0 hours, full fine-tuning still yields better results. However, at the 2.0-hour mark, Low-Rank Adaptation successfully overtakes full fine-tuning, achieving a WER of 75.66 compared to 84.10. Therefore, these empirical findings indicate that parameter-efficient methods can effectively prevent model overfitting when adapting to highly distant languages.

\begin{table*}[tbp]
\centering
\caption{Evaluation of Catastrophic Forgetting on the English LibriSpeech Test Set. The table illustrates the degradation in English transcription accuracy after the Whisper-Small model is adapted to the extreme low-resource Lelepa dataset.}
\label{tab:forgetting_results}
\scalebox{0.85}{
\begin{tabular}{llcccc}
\toprule
\textbf{Model} & \textbf{Adaptation Strategy} & \textbf{Updated Params} & \textbf{Chinese Test CER} & \textbf{English Test WER} & \textbf{French Test WER}\\
\midrule
whisper-small & None (Before Fine-Tuning) & 0 & 34.62 & 15.68 & 24.27 \\
whisper-small & Low-Rank Adaptation (LoRA) & 20.1M & 36.71 & 18.89 & 27.82    \\
whisper-small & Full Fine-Tuning (FullFT) & 244M & 45.02 & 26.24 & 40.26 \\
\bottomrule
\end{tabular}}
\vspace{-4mm}
\end{table*}

\begin{table}[tbp]
\centering
\caption{Impact of Component-Specific Adaptation on Target Accuracy and Catastrophic Forgetting. We evaluate Whisper-Small fine-tuned on Lelepa and test its knowledge retention on English. The results reveal a strict trade-off between target acoustic adaptation and source language retention.}
\label{tab:module_ablation}
\scalebox{0.7}{
\begin{tabular}{llcc}
\toprule
\textbf{Adaptation} & \textbf{Updated} & \textbf{Lelepa CER ($\downarrow$)} & \textbf{English WER ($\downarrow$)} \\
\midrule
Baseline         & None                     & - & 15.68 \\
FullFT    & Encoder + Decoder        & 27.27  & 26.24 \\
\midrule
LoRA            & Encoder + Decoder        & 28.41  & 21.24 \\
Decoder-only LoRA            & Decoder Only             & 34.68  & 18.27 \\
Encoder-only LoRA            & Encoder Only             & 29.87  & 31.26 \\
\bottomrule
\end{tabular}}
\vspace{-4mm}
\end{table}

\subsection{Layer-Wise Representational Drift Analysis}

To uncover \textit{why} these performance divergences occur, we open the black box to analyze the layer-wise representational drift before and after adaptation.
Specifically, we compare the hidden states of the base model and the model fine-tuned for 2.0 hours across all 12 layers. 
First, the encoder drift patterns show a clear divergence based on language difficulty. For Bislama and Nafsan, the internal changes primarily occur in the later encoder layers. This pattern suggests that the model reuses the basic acoustic features from early layers and only updates the high-level phonetic representations.

In contrast, Lelepa exhibits significant drift in the early encoder layers. This early shift indicates that the basic acoustic properties of Lelepa differ greatly from the pretraining data. Therefore, the model must reconstruct foundational features. Furthermore, the decoder drift highlights this difference. Bislama and Nafsan show active adjustments across most middle and late decoder layers. However, Lelepa displays almost no drift between layers 2 and 10, with a sudden peak at the final layer. Consequently, these layer-wise behaviors directly explain the adaptation performance. Because Lelepa requires deep acoustic restructuring, standard full fine-tuning struggles. This observation logically explains why parameter-efficient methods like LoRA achieve lower error rates than full fine-tuning at the 2.0-hour mark for Lelepa, as detailed in Table 3. 

\subsection{Quantifying Catastrophic Forgetting}
To rigorously evaluate the model’s plasticity–stability trade-off, we quantify catastrophic forgetting after adaptation. Table \ref{tab:forgetting_results} reports error rates for high-resource languages that were already included during pre-training and shows that forgetting clearly emerges for these seen languages after adapted to the Pacific languages. Updating all parameters severely distorts the pre-trained multilingual representations, leading to substantially stronger forgetting than with LoRA-based updating. 
Although LoRA approach preserves knowledge more effectively than full fine-tuning, a noticeable loss in accuracy remains.

Therefore, we systematically isolate the parameter updates to specific architectural modules (i.e., updating encoder, decoder, or both) to find the exact source of this forgetting, as detailed in Table \ref{tab:module_ablation}. This architectural dissection reveals a strict trade-off between target acoustic adaptation and source language retention. Restricting adaptation exclusively to the decoder minimizes forgetting, yielding an English WER of 18.27 percent. However, this configuration fails to learn the target task, resulting in a high CER of 34.68 percent for Lelepa. Conversely, applying adaptation exclusively to the acoustic encoder improves the target recognition to 29.87\%. Yet, this acoustic focus causes severe catastrophic forgetting, pushing the English error rate to 31.26 percent. Surprisingly, this degradation is worse than the full parameter updating approach. Consequently, these findings indicate that adapting the entire encoder destroys universal acoustic features, while restricting updates to the decoder prevents effective target learning. Therefore, a simple binary choice between the encoder and decoder is insufficient for unseen pacific languages.

\begin{table}[tbp]
\centering
\caption{Sequential Continual Learning Performance: Comparison of Full FT and LoRA on Nafsan $\rightarrow$ Lelepa Sequence (Whisper-Small)}
\vspace{-3mm}
\label{tab:seqcl_matrix_updated}
\resizebox{0.9\linewidth}{!}{
\begin{tabular}{llccc}
\toprule
\multirow{2}{*}{\textbf{Strategy}} & \multirow{2}{*}{\textbf{Training Stage}} & \multicolumn{2}{c}{\textbf{Test WER (\%)}} &  \multirow{2}{*}{\textbf{Avg. WER}} \\ \cmidrule(lr){3-4}
&  & \textbf{Nafsan} & \textbf{Lelepa}   &  \\ \midrule
\multirow{2}{*}{Full FT} & After Nafsan & 47.02 & N/A  & \multirow{2}{*}{\textbf{64.70\%}} \\
& After Lelepa & 45.67 & 83.72   &  \\ \midrule
\multirow{2}{*}{LoRA~\cite{hu2022lora}} & After Nafsan & 53.23 & N/A & \multirow{2}{*}{76.52\%} \\
& After Lelepa & 84.42 & 68.62   &  \\ \midrule
\multirow{2}{*}{DoRA~\cite{dora}} & After Nafsan & 53.57 & N/A  & \multirow{2}{*}{77.81\%} \\
& After Lelepa & 84.80 & 70.82   &  \\ \midrule
\multirow{2}{*}{O-LoRA~\cite{wang2023orthogonal}} & After Nafsan & 53.47 & N/A  & \multirow{2}{*}{78.25\%} \\
& After Lelepa & 87.42 & 69.08    &  \\ \bottomrule
\end{tabular}}
\vspace{-4mm}
\end{table}

\subsection{Sequential training and the Forgetting}
We further quantify how catastrophic forgetting emerges when the model sequentially learns multiple low‑resource languages. In this continual learning scenario, we first train the models on the Nafsan corpus. Subsequently, we fine-tune them on the Lelepa dataset. We then evaluate their immediate plasticity on the new task alongside their historical stability on the previously learned language. 
In addition to LoRA, we also evaluate two more regularized variants: Weight-Decomposed Low-Rank Adaptation (DoRA)~\cite{dora} which separates weight magnitude and direction, and ii) Orthogonal Low-Rank Adaptation (O-LoRA)~\cite{wang2023orthogonal} which enforces independent training stages. We test whether these methods protect historical knowledge.

As shown in Table \ref{tab:seqcl_matrix_updated}, full Fine-Tuning shows excellent historical stability. After learning Lelepa, its error rate on Nafsan remains surprisingly low at 45.67 percent. However, it severely struggles to learn the new language, yielding a high error rate of 83.72 percent on Lelepa. In contrast, the parameter-efficient methods demonstrate better immediate plasticity on the new task. For example, standard LoRA achieves a lower error rate of 68.62 percent on Lelepa. Furthermore, DoRA and O-LoRA show similar improvements on the new target. Nevertheless, these efficient methods suffer from severe catastrophic forgetting. Their error rates on the previously learned Nafsan jump significantly to over 84 percent. Consequently, Full Fine-Tuning maintains the lowest average WER of 64.70 percent. Therefore, these results reveal a critical plasticity and stability dilemma. While parameter-efficient methods adapt well to new domains, they fail to protect the historical knowledge during sequential learning for pacific languages.

\section{Discussion and Conclusion}
Synthesizing our results establishes guidelines for deploying speech models in Pacific communities. First, we contradict the assumption of universal adaptability. Adapting to low-resource languages like Lelepa and Nafsan causes severe catastrophic forgetting. This failure occurs because their vast linguistic distance forces models to overwrite original representations. Second, extreme data imbalance complicates this adaptation. For example, Bislama possesses significantly more data than Nafsan, which forces models to overfit. Consequently, Full Fine-Tuningis often suboptimal for these scenarios. Furthermore, sequential evaluations confirm that LoRA provides immediate plasticity but fails to protect historical knowledge. Therefore, current methods leave the stability and plasticity dilemma unresolved. In conclusion, studying Pacific Indigenous languages exposes the structural vulnerabilities of standard methods. A simple binary choice between updating the encoder or the decoder is insufficient. Therefore, future research must explore dynamic architectures~\cite{xiao2025analytickws, chen2025aft}. Developing new methods to handle unique linguistic features and unbalanced data remains a critical direction.

\clearpage

\section{Generative AI Use Disclosure}
We use generative AI tools for polishing the manuscript, e.g., correcting the grammar.

\bibliographystyle{IEEEtran}
\bibliography{mybib}

\end{document}